\documentclass[a4paper, 12pt]{article}%{revtex4}
\usepackage[cp1251]{inputenc} %- для WiN
\usepackage[english, russian]{babel} %- для рус-англ. переноса
\usepackage{amsmath, amssymb,bm, amsfonts, graphics, latexsym, array}
\usepackage{cite}
\usepackage[dvips]{graphicx}
\usepackage[flushleft,small]{caption2} % чтобы в подписях к рисункам

\usepackage[dvips]{graphicx}

\begin{document}
\large
\renewcommand{\abstractname}{\,Abstract}
\renewcommand{\refname}{\begin{center} REFERENCES\end{center}}

 \begin{center}\large
\bf Chemical Potential Jump during Evaporation of a
Quantum Bose Gas
\end{center}\medskip
\begin{center}
\bf
  E. A. Bedrikova\footnote{$bedrikova@mail.ru$}
  A. V. Latyshev\footnote{$avlatyshev@mail.ru$}
\end{center}\medskip

\begin{center}
{\it Faculty of Physics and Mathematics,\\ Moscow State Regional
University, 105005,\\ Moscow, Radio str., 10--A}
\end{center}\medskip

\thispagestyle{empty}
\large

\begin{abstract}
The dependence of the chemical potential jump coefficient on the
evaporation coefficient is analyzed for the case in which the
evaporating component is a Bose gas. The concentration of the
evaporating component is assumed to be much lower than the
concentration of the carrier gas. The expression for the chemical
potential jump is derived from the analytic solution of the
problem for the case in which the collision frequency of molecules
of the evaporating component is constant.\\

 Jp Boson systems;
71.35.Lk   Collective effects
(Bose effects, phase space filling, and excitonic phase
transitions).

\end{abstract}

\begin{center}
{\bf INTRODUCTION}
\end{center}

Interest in problems associated with the behavior of gases under
conditions in which quantum properties cannot be ignored has been
growing in recent years \cite{1}. The behavior of mixtures of such gases
is also of considerable interest. The "quantum"\, gases that are
studied most frequently are  $^3$He  и $^4$He. It should be noted that
$^3$He is a Fermi gas, while $^4$He is a Bose gas. Such a combination
of different quantum-mechanical statistics attracts considerable
attention to the mixtures of these gases \cite{2}.
A large number of publications are also devoted to the properties
of solutions of these gases \cite{3,4}. At the same time, boundary
value problems for such mixtures have been studied insufficiently.
Such problems include the behavior of a mixture of quantum gases
in the vicinity of the evaporation boundary. Let us consider
the most frequently encountered case of a dilute mixture.
Let molecular concentration $n_1$ of one gas be much lower than
molecular concentration $n_2$ of the other gas: $n_1 \ll n_2$. We will
consider the problem of evaporation of the first gas to the gas
mixture.
We will henceforth consider the problem in a more general
formulation, taking into account the possibility of other
applications also. Let us suppose that we have a mixture of
two quantum gases. We consider the problem of evaporation of
a Bose gas to this mixture, provided that its concentration
is much lower than the concentration of the other component.

\begin{center}
  \bf 1. FORMULATION OF THE PROBLEM AND BASIC EQUATIONS
\end{center}

Let us consider evaporation from a plane surface to a binary gas
mixture. We assume that concentration $n_1$
of the evaporating component of the mixture is much lower than
concentration $n_2$ of the nonevaporating component
($n_1\ll n_2$; dilute mixture). It should be noted that this condition
holds in the most important applications.

In the semiclassical approximation, the Boltzmann equation for a
binary gas mixture has the form \cite{5}
$$
\dfrac{\partial f_i}{\partial t}+{\mathbf{v}}_i
\dfrac{\partial f_i}{\partial {\mathbf{r}}_i}=J_{ii}+J_{ij},
\qquad\; i\ne j, \; i,j=1,2.
\eqno{(1)}
$$

Here, $f_i$ is the distribution function for the $i$th component
of the mixture and $J_{ii}$ and $J_{ij}$ are the integrals of collisions
of molecules of the $i$th component with
one another and with molecules of the $j$th component, respectively.

It should be noted that
$J_{11}\sim n_1^2$ and $J_{12}\sim n_1 n_2$ (since
$f_i\sim n_i$).
Quantity $\varepsilon=n_1/n_2$ is a small parameter
($\varepsilon \ll 1$), since $n_1\ll n_2$.
Obviously, $|J_{11}|/|J_{12}|\sim\varepsilon$.
Consequently, quantity $J_{11}$ can be disregarded as compared to $J_{12}$ in
the first approximation in $\varepsilon$. In addition, the action of the first
component on the distribution function for the second component can
also be disregarded in this approximation in $\varepsilon$. Therefore, under the
conditions of the given problem, the distribution function for the
second component of the gas mixture can be treated as an equilibrium
function with mean velocity ${\bm U}_2=0$ and constant tempera\-tu\-re $T$ and
concentration  $n_2$.

Quantity $J_{12}$ can be approximated by a kinetic model of the
Bhat\-na\-gar---Gross---Krook (BGK) type \cite{6,7}. Taking into account
Eq. (1), we can write the kinetic equation for the first component in the form
$$
\dfrac{\partial f}{\partial t}+{\mathbf{v}}_1
\dfrac{\partial f}{\partial {\mathbf{r}}}=\nu_1
 \left(f_{eq}-f\right),
\eqno{(2)}
$$
where
$$
 f_{eq}=\Big[-1+\exp\Big(\dfrac{mv^2}{2kT}-
 \dfrac{\mu(\mathbf{r})}{kT}\Big)\Big]^{-1},\qquad \mu(\mathbf{r})<0,
$$
$f_{eq}$ is the equilibrium Bose distribution function for the first
component, $k$ is the Boltzmann constant,  $\nu_1$ is the effective collision
frequency for molecules of the first component, $\mu(\mathbf{r})$ is
the chemical potential of the first component, and $T$ is
the temperature of the mixture, which is assumed to be constant.

Let us choose the Cartesian system of coordinates with the center
on the surface from which evaporation takes place. We direct the $x_1$
axis along the normal to the surface. During evaporation from the surface,
a constant gradient of the first component concentration
$$
g_n=\Big(\frac{d\,n}{dx_1}\Big)_{x_1=\infty}
$$
exists at a large distance from the surface.

Here
$$
n(x_1)=\int f(x_1,\mathbf{v})\,d\Omega, \qquad
d\Omega=\dfrac{(2s+1)\,d^3p}{(2\pi \hbar)^3}.
$$
is the concentration (number density) of the first component
and $\mathbf{p}$ is the momentum of molecules of the first component.

We assume that evaporation is weak; in other words, we assume
that the relative variation in the concentration of the first component over
the mean free path $l$ of molecules is much smaller than unity:
$$
|G_n|\ll 1,\qquad \;G_n=\frac{l}{n_{s}}|g_n|.
$$

Here, $n_{s}$ is the concentration of saturated vapor (gas) of the first
component on the evaporation
surface, which corresponds to surface temperature $T_s\equiv T$.

Under such conditions, the problem can be linearized.
Preliminarily, we pass to dimensionless velocity $\mathbf{C}=\mathbf{v}/v_T$,
dimensionless coordinate $x=x_1/l_T$ and
dimensionless chemical potential $\alpha(x)=\mu(x)/(kT)$.

Here $v_T=1/\sqrt{\beta}$ is the thermal velocity of the first component,\;
$\beta=m/(2kT)$, $l=v_T\tau$ is the mean free thermal path of molecules
of first component, $\tau=1/\nu_1$ is the mean time between two
collisions of molecules of the first component.

We can now write Eq. (2) in the form
$$
C_x\dfrac{\partial f}{\partial x}=f_F(x,C)-f(x,\mathbf{C}).
\eqno{(3)}
$$

Here $f_B(x,C)$ is a locally equilibrium Bose distribution function,
$$
f_B(x,C)=\Big[-1+\exp(C^2-\alpha(x))\Big]^{-1}.
$$

We will linearize the problem relative to the absolute Bosean
$$
f_B^s\equiv f_B^s(C,\alpha_s)=\Big[-1+\exp(C^2-\alpha_s)\Big]^{-1},
$$
where $\alpha_s$ is the value of the dimensional
chemical potential corresponding to surface temperature $T \equiv T_s$
and the concentration of the saturated vapor at this temperature.

We will linearize the dimensionless chemical potential relative to
quantity $\alpha_s$; i.e., we assume that $\alpha(x)=\alpha_s+a(x)$.
Linearizing the locally equilibrium Bose
distribution function relative to $f_B^s$, we obtain
$$
f_B^s(x,C)=f_B^s(C,\alpha_s)+g(C,\alpha_s)a(x),
$$
where
$$
g(C,\alpha_s)=\exp(C^2-\alpha_s)(-1+\exp(C^2-\alpha_s))^{-2}.
$$

We will seek the distribution function in the form
$$
f(x,\mathbf{C})=f_F^s(C,\alpha_s)+
g(C,\alpha_s)h(x,C_x).
\eqno{(4)}
$$

Using this expression (4), we can write Eq. (3) as:
$$
C_x\dfrac{\partial h}{\partial x}=a(x)-h(x,C_x).
\eqno{(5)}
$$

We can now find the dimensionless relative deviation
$a(x)=\alpha(x)-\alpha_s$. of the chemical potential from its value at the
wall using the law of conservation of the number of particles:
$$
\int f(x,\mathbf{C})\;d\Omega=\int f_B(x,C)d\Omega.
$$

The law of conservation of the number of particles leads to the equation
$$
\int [h(x,\mathbf{C})-a(x)]g(C,\alpha_s)\,d^3C=0,
$$
which gives
$$
a(x)=\dfrac{\displaystyle\int h(x,C_x)g(C,\alpha_s)\,d^3C}
{\displaystyle\int g(C,\alpha_s)\,d^3C}.
\eqno{(6)}
$$

We can easily find that
$$
\int g(C,\alpha_s)\,d^3C=
4\pi \int\limits_{0}^{\infty}g(C,\alpha_s)C^2\,dC=
2\pi f_0(\alpha_s),
$$
where
$$
f_0(\alpha_s)=\int\limits_{0}^{\infty}f_B(C,\alpha_s)\,dC=
\int\limits_{0}^{\infty}\dfrac{dC}{-1+\exp(C^2-\alpha_s)}.
$$

We transform the numerator of expression (6) as follows:
$$
\int h(x,C_x)g(C,\alpha_s)\,d^3C
=\pi\int\limits_{-\infty}^{\infty}f_B(C_x,\alpha_s)
h(x,C_x)\,dC_x.
$$

Consequently, in accordance with relation (6), the
deviation of the dimensionless chemical potential is given by
$$
a(x)=\dfrac{1}{2f_0(\alpha_s)}\int\limits_{-\infty}^{\infty}
f_B({\mu},\alpha_s)h(x,\mu)\,d\mu.
\eqno{(7)}
$$

Thus, kinetic equation (5) for the problem of evaporation,
taking into account relation (7), has the form
$$
\mu\dfrac{\partial h}{\partial x}+h(x,\mu)=\dfrac{1}{2f_0(\alpha_s)}
\int\limits_{-\infty}^{\infty}f_B({\mu'},\alpha_s)h(x,\mu')\,d\mu',\quad
\mu=C_x.
\eqno{(8)}
$$\\

\begin{center}
  \bf 2. FORMULATION OF THE BOUNDARY CONDITIONS
\end{center}

Let us consider the boundary condition at the evaporation surface
for molecules of the first component taking into account the effect of the
surface by introducing evaporation coefficient $q$ (see \cite{8,9}),
$$
f(0,{\mathbf{C}})=q f_B^s(C, \alpha_s) +
(1-q)f_B^0(C,\alpha_0),\quad C_{x}>0,
\eqno{(9)}
$$
where
$$
f_B^0(C,\alpha_0)=\Big[-1+\exp(C^2-\alpha_0)\Big]^{-1}.
$$

Quantity  $\alpha_0$ can be determined from the nonpercolation conditions
for molecules reflected from the surface without being
condensed on it (the probability of such a process is $1-q$),
$$
(1-q)\int C_{x}[f_B^0(C,\alpha_0)\theta_+(C_x) +
f(0,\mathbf{C})\theta_+(-C_x)]d\Omega=0,
$$
where $\theta_+(x)$ ) is the Heaviside function, $\theta_+(x)=1$ for
$x>0$ and $\theta_+(x)=0$ for $x<0$.

Taking into account the definition of the Heaviside function,
we can transform the nonpercolation condition to
$$
\int\limits_{C_x>0} C_xf_B^0(C,\alpha_0)\,d^3C+
\int\limits_{C_x<0} C_xf(0,\mathbf{C})\,d^3C=0.
\eqno{(10)}
$$

The boundary condition for function $h(x,C_x)$ at the wall can
be derived from condition (10). Substituting function (4) into (10),
taking into account
condition (9) as well as the results of the linearization
$$
f_B^0(C,\alpha_0)=\dfrac{1}{-1+\exp(C^2-\alpha_0)}=
\dfrac{1}{-1+\exp(C^2-\alpha_s-(\alpha_0-\alpha_s))}=
$$
$$
=f_B^s(C,\alpha_s)+g(C,\alpha_s)(\alpha_0-\alpha_s).
$$
and carrying out the substitution,
we obtain the following boundary condition at the wall:
$$
h(0,\mu)=(1-q)(\alpha_0-\alpha_s),\qquad \mu>0.
\eqno{(11)}
$$

At a large distance from the surface (outside the Knudsen
layer having a thickness on the order of
the mean free path of molecules), function  $h(x,C_x)$ has the form
$$
h(x,\mu)=h_{as}(x,\mu)+o(1),\quad C_x=\mu, \quad x\to +\infty,
\eqno{(12)}
$$
where
$$
h_{as}(x,\mu)=A_\alpha+G_\alpha(x-\mu),
$$
$G_\alpha$ here, $G_\alpha$ is the gradient of the
dimensionless chemical potential defined far away from the wall,
$$
G_\alpha=\Big(\dfrac{d\,\alpha(x)}{d\,x}\Big)_{x=+\infty},
$$
and $A_\alpha$  is the chemical potential jump, viz.,
the unknown quantity to be determined from the solution of the problem.

Substituting function $h_{as}(x,\mu)$ into definition (6) of the
dimensional chemical potential, we obtain the asymptotic
distribution of this potential:
$$
a_{as}(x)=A_\alpha+G_\alpha x, \qquad x\to +\infty.
$$

It follows, hence, that $A_\alpha=a_{as}(0)=\alpha_{as}(0)-\alpha_s$,
i.e., the chemical potential jump is defined as the difference
between the extrapolated value of
chemical potential at the wall and its value immediately at the wall.

Quantity $\alpha_0-\alpha_s$ can be determined from nonpercolation condition (10).
This condition can be written in the explicit form
$$
\int\limits_{C_x>0}C_x\Big[f_B^s(C,\alpha_s)+g(C,\alpha_s)
(\alpha_0-\alpha_s)\Big]\,d^3C+
$$
$$
+\int\limits_{C_x<0}C_x\Big[f_B^s(C,\alpha_s)+g(C,\alpha_s)
h(0,C_x)\Big]\,d^3C=0.
$$

It should be noted that the sum of the integrals of the first
and second terms in each square bracket is zero;
therefore, we arrive at the following equation:
$$
(\alpha_0-\alpha_s)\int\limits_{C_x>0}C_x\,g(C,\alpha_s)\,d^3C+
\int\limits_{C_x<0}C_x\,g(C,\alpha_s)h(0,C_x)\,d^3C=0.
$$

The second integral, which can be evaluated over the negative
half-space, will be calculated using the conservation law for
momentum (to be more precise, the $x$ component of the momentum).
For this purpose, we replace $h(0,C_x)$ from the first
integral on the right-hand side of the obvious equality
$$
\int\limits_{C_x<0}C_x\,g(C,\alpha_s)h(0,C_x)\,d^3C=
$$
$$
=\int\,C_x\,g(C,\alpha_s)h(0,C_x)\,d^3C-
\int\limits_{C_x>0}C_x\,g(C,\alpha_s)h(0,C_x)\,d^3C
$$
by $h_{as}(0,C_x)$ and $h(0,C_x)$ from the second integral by
$(1-q)(\alpha_0-\alpha_s)$ from boundary condition (11). After evaluating the
required integrals, we arrive at the equation, which gives
$$
\alpha_0-\alpha_s=\dfrac{4g_4(\alpha_s)}{3g_3(\alpha_s)}\cdot
\dfrac{G_\alpha}{q}.
\eqno{(13)}
$$

Here
$$
g_{n+2}(\alpha_s)=\int\limits_{0}^{\infty}C^{\,n+2}
g(C,\alpha_s)\,dC,\quad n=0,1,2.
$$

Evaluating the first two of these integrals, we obtain
$$
g_3(\alpha_s)=-\dfrac{1}{2}\ln(1-e^{\alpha_s}), \qquad
g_4(\alpha_s)=\dfrac{3}{4}l(\alpha_s),
$$
$$
l(\alpha_s)=-\int\limits_{0}^{\infty}\ln(1-e^{\alpha_s-C^2})\,dC.
$$

Consequently, equality (13) can be written in the explicit form
$$
\alpha_0-\alpha_s=-\dfrac{2l(\alpha_s)}{\ln(1-e^{\alpha_s})}\cdot
\dfrac{G_\alpha}{q}.
$$

Now, boundary condition (11) is complete
$$
h(0,\mu)=B,\quad \mu>0,\quad \text{где}\quad
B=-\dfrac{1-q}{q}\cdot \dfrac{2l(\alpha_s)}{\ln(1-e^{\alpha_s})}
G_{\alpha}.
\eqno{(14)}
$$

Thus, the boundary value problem involves the obtaining a solution
to Eq. (8), which satisfies boundary conditions
(14) and (12). Quantity $\alpha_s$ will be henceforth denoted as $\alpha$.
\\

\begin{center}
  \bf 3.  DIFFUSION COEFFICIENT AND THE MASS FLOW RATE
\end{center}

Function
$$
f_{as}(x,\mathbf{C})=f_B^s(C,\alpha)+
g(C,\alpha)h_{as}(x,C_x),
$$
where
$h_{as}(x,\mu)=A_\alpha+G_\alpha(x-\mu)$,
is known as the Chapman---Enskog distribution function \cite{7,8}.

Using this function, we can calculate diffusion coefficient $D_{12}$.
Diffusion flux $\mathbf{i}$ emerges due to the presence of a density gradient
$\nabla \rho$ in the gas; consequently, for small values of
$\nabla \rho$, we have $\mathbf{i}=-D_{12}\nabla \rho$  \cite{10}.
In accordance with the formulation of the problem, this gives
$$
D_{12}=-\dfrac{i_x}{\rho'_{x_1}}=-
\dfrac{m\int f_{as}v_x\,d\Omega}{m\int (f_{as})'_{x_1}\,d\Omega}.
$$

Here, we replace dimensionless coordinate $x$ by
dimensional coordinate $x_1=x/(\nu_1 \sqrt{\beta})$ and pass to integration with
respect to dimensionless velocity. This gives
$$
D_{12}=\dfrac{4kT}{3m\nu_1}\cdot \dfrac{g_4(\alpha)}
{f_0(\alpha)}.
\eqno{(15)}
$$

It should be noted that this result is transformed into the classical
one for $\alpha\to -\infty$ \cite{9}. Indeed, using the asymptotic form
$$
g_4(\alpha)=e^\alpha \dfrac{3\sqrt{\pi}}{8}, \quad
f_0(\alpha)=e^\alpha \dfrac{\sqrt{\pi}}{2}, \quad \alpha\to -\infty,
$$
we obtain the known result
$$
D_{12}=\dfrac{kT}{m\nu_1}.
$$

It should be noted that, in the approach used here, diffusion
coefficient $D_{12}$  is treated as an empirical quantity. For example,
for the  $^3$He -- $^4$He mixture at a temperature of $2 K$, we have \cite{He}
$D_{12}\simeq 10^{-3}$ см$^2$/с.

Let us find the mass flow rate of the evaporating component
of the binary gas in the direction of the $x$ axis.
By definition, the mass flow rate in the $x$ direction is
$$
U_x=\dfrac{1}{N}\int fv_x\,d\Omega, \qquad N=\int f\,d\Omega.
$$

Substituting this expression into the definition of
diffusion coefficient, we obtain
$$
D_{12}=-\dfrac{\displaystyle\int f_{as}v_x\,d\Omega}{N}\cdot
\dfrac{N}{\displaystyle\int (f_{as})'_{x_1}\,d\Omega}=U_x\cdot
\dfrac{N}{\displaystyle\int (f_{as})'_{x_1}\,d\Omega}.
$$

It can easily be seen that
$$
D_{12}=U_x\cdot N \Big[\nu_1\sqrt{\beta}
\int g(C,\alpha)\,d\Omega\Big]^{-1}G_\alpha^{-1},
$$
passing to dimensionless mass flow rate $W_x=\sqrt{\beta}U_x$, we obtain
$$
D_{12}=\dfrac{W_x}{G_\alpha}\cdot
\dfrac{2kT}{\nu_1m}\cdot \dfrac{\pi^2N}{(2s+1)k_T^3g_2(\alpha)},
$$
where
$$
f_0(\alpha)=\int\limits_{0}^{\infty}f_B(C,\alpha)\,dC,\qquad
k_T=\dfrac{mv_T}{\hbar},
$$
where $k_T$ is the thermal wave number.

The number density in the linear approximation is given by
$$
N=\int f_B(C,\alpha)\,d\Omega=\dfrac{(2s+1)k_T^3}{2\pi^2}f_2(\alpha),
$$
where
$$
f_2(\alpha)=\int\limits_{0}^{\infty}f_B(C,\alpha)C^2dC=
\int\limits_{0}^{\infty}\dfrac{C^2dC}{-1+e^{C^2-\alpha}}=\dfrac{1}{2}l(\alpha).
$$

Thus, we obtain
$$
D_{12}=\dfrac{W_x}{G_\alpha}\cdot
\dfrac{2kT}{\nu_1m}\cdot\dfrac{f_2(\alpha)}{g_2(\alpha)},
$$
whence the mass flow rate is
$$
W_x=\dfrac{\nu_1m}{2kT}D_{12}
\dfrac{f_0(\alpha)}{l(\alpha)}G_\alpha.
$$\\

\begin{center}
  \bf 4. SOLUTION OF THE PROBLEM
\end{center}

We will seek the solution to Eq. (8) in the form
$$
h_\eta(x,\mu)=\exp(-\dfrac{x}{\eta})\Phi(\eta,\mu),
$$
$$
\dfrac{1}{2f_0(\alpha)}
\int\limits_{-\infty}^{\infty}f_B(\mu,\alpha)
\Phi(\eta,\mu)\,d\mu=1,
$$
where $\eta$ ц is the spectral parameter or the separation parameter.

Using these two equalities, we obtain from Eq. (8) the
characteristic equation
$$
(\eta-\mu)\Phi(\eta,\mu)=\eta.
$$
For $\eta\in (-\infty,\infty)$
we obtain the eigenfunctions of the characteristic equation,
$$
\Phi(\eta,\mu)=\eta P\dfrac{1}{\eta-\mu}+
\dfrac{2f_0(\alpha)}{f_B(\eta,\alpha)}
\lambda(\eta)\delta(\eta-\mu).
$$

Here, the symbol  $P x^{-1}$ indicates the principal value of the
integral of $x^{-1}$, $\delta(x)$ is the Dirac delta
function, and $\lambda(z)$  is the dispersion function of the problem,
$$
\lambda(\eta)=1+\dfrac{\eta}{2f_0(\alpha)}
\int\limits_{-\infty}^{\infty}
\dfrac{f_B(\mu,\alpha)d\mu}{\mu-\eta}.
$$

\begin{figure}[h]
\begin{center}
\includegraphics[width=14cm,height=9cm]{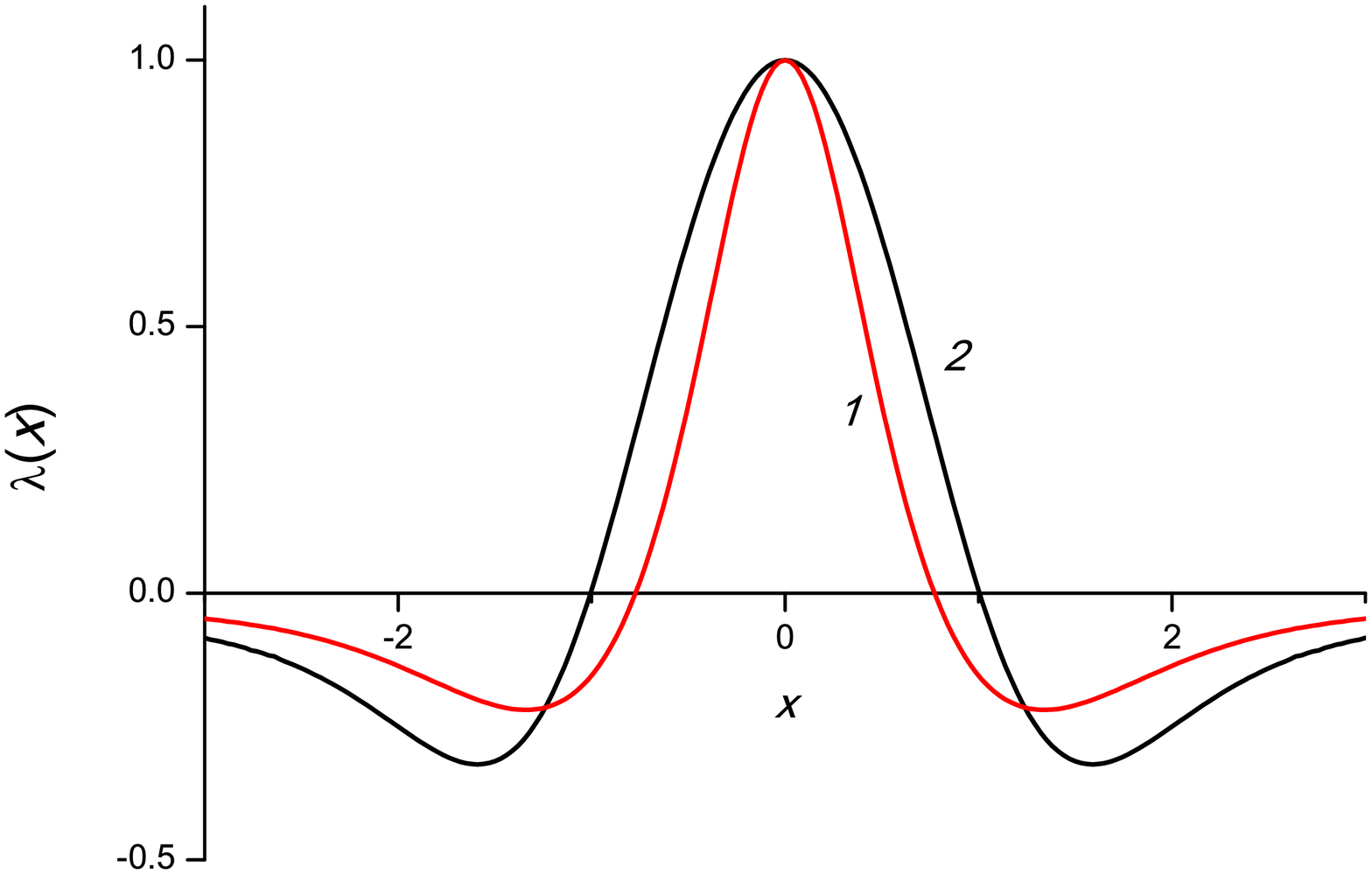}
\end{center}
\begin{center}
\caption{Dispersion function on real axis, curve $1$ correspond to Bose--gases,
curve $2$ correspond to Fermi--gases from \cite{Kost}, $\alpha=-1$.}
\end{center}
\end{figure}

We will seek the solution to problem (8), (12), and (14)
in the form of an expansion
$$
h(x,\mu)=A_\alpha+G_\alpha(x-\mu)+\int\limits_{0}^{\infty}
\exp(-\dfrac{x}{\eta})\Phi(\eta,\mu)A(\eta)\,d\eta,
\eqno{(16)}
$$
where $A(\eta)$ is an unknown function (coefficient of the continuous
spect\-rum) and $A_\alpha$  is an unknown constant (coefficient of the discrete
spect\-rum). It will be shown below that coefficients
$A_\alpha$ and $G_\alpha $ are connected by a linear relation,
$$
A_\alpha=C(\alpha,q)G_\alpha,
\eqno{(17)}
$$
in which function $C(\alpha,q)$ is the coefficient of the chemical
potential jump.

Let us find the coefficients of the discrete and continuous spectra.
The substitution of eigenfunctions into expansion (16) and of
expansion (16) into boundary condition (14) leads to a singular
integral equation with the Cauchy kernel:

$$
B=A_\alpha-G_\alpha\mu+\int\limits_{0}^{\infty} \dfrac{\eta
A(\eta)\,d\eta}{\eta-\mu}\,d\eta+\dfrac{2f_0(\alpha)}
{f_B(\mu,\alpha)}\lambda(\mu)A(\mu)=0,\;\mu>0,
$$

We introduce an auxiliary function,
$$
M(z)=\int\limits_{0}^{1}
\dfrac{\eta A(\eta)\,d\eta}{\eta-z},
$$
with upper and lower boundary values on the positive semiaxis
related by the Sokhotsky formulas
$$
M^+(\mu)-M^-(\mu)=2\pi i \mu A(\mu), \qquad \mu>0,
\eqno{(18)}
$$
$$
\dfrac{1}{2}[M^+(\mu)+M^-(\mu)]=M(\mu),\qquad
M(\mu)=\int\limits_{0}^{1}
\dfrac{\eta A(\eta)\,d\eta}{\eta-\mu},
$$
and the last integral in which is treated as a singular integral
in the sense of the Cauchy principal value.

Using the upper and lower boundary values of functions  $\lambda(z)$ and
$M(z)$ on the cut $(0,\infty)$, we can reduce the singular equation to the
Riemann boundary value problem:
$$
\lambda^+(\mu)[M^+(\mu)+A_\alpha-G_\alpha\mu-B]=
$$
$$=
\lambda^-(\mu)[M^-(\mu)+A_\alpha-G_\alpha\mu-B],\;
\quad \mu>0.
$$

Let us consider the corresponding homogeneous boundary value prob\-lem
$$
\dfrac{X^+(\mu)}{X^-(\mu)}=\dfrac{\lambda^+(\mu)}
{\lambda^-(\mu)}, \quad \mu>0.
$$

Its solution \cite{12} has the form
$$
X(z)=\dfrac{1}{z}\exp V(z), \quad
V(z)=\dfrac{1}{\pi}\int\limits_{0}^{\infty}
\dfrac{\zeta(\tau,\alpha)\,d\tau}{\tau-z},
$$
where
$$
\zeta(\tau,\alpha)=\arcctg\dfrac{2f_0(\alpha)\lambda(\tau)}
{\pi\tau f_B(\tau,\alpha)}-\pi.
$$

Using the homogeneous problem, we can reduce the inhomogeneous
problem to determining
the analytic function from its zero-point jump at the cut:
$$
X^+(\mu)[M^+(\mu)+A_\alpha-G_\alpha\mu-B]=$$$$
=X^-(\mu)[M^-(\mu)+A_\alpha-G_\alpha\mu-B],\;\quad \mu>0.
$$

The solution to this problem has the form
$$
M(z)=-A_\alpha+G_\alpha z+B-\dfrac{G_\alpha}{X(z)}.
$$

Using the condition $M(\infty)=0$, we obtain
the jump of the dimen\-sion\-less chemical potential,
$$
A_\alpha=V_1(\alpha)G_\alpha+B,\quad\text{where}\quad
V_1(\alpha)=-\dfrac{1}{\pi}\int\limits_{0}^{\infty}
\zeta(\tau,\alpha)\,d\tau.
\eqno{(19)}
$$

Coefficient $A(\eta)$ of the continuous spectrum can be determined
by substituting the
solution to the boundary value problem into equality (18):
$$
A(\eta)=-\dfrac{G_\alpha}{2\pi i \eta}\Big[\dfrac{1}{X^+(\eta)}-
\dfrac{1}{X^-(\eta)}\Big]=\dfrac{\sin\zeta(\eta)}{\pi \eta X(\eta)}G_\alpha.
$$

Thus, the unknown coefficients in expansion (16) are determined.
This means that the distribution func-tion for the evaporating component
is constructed completely and can be written in the form
$$
\dfrac{h(x,\mu)}{G_\alpha}=C(\alpha,q)+x-\mu+\dfrac{1}{\pi}
\int\limits_{0}^{\infty}
\exp\Big(-\dfrac{x}{\eta}\Big)\dfrac{\Phi(\eta,\mu)}{\eta X(\eta)}
\sin\zeta(\eta)d\eta.
$$

Using the contour integration methods for boundary  $x=0$, we obtain
$$
\dfrac{h(0,\mu)}{G_\alpha}=
B-\dfrac{\cos \zeta(\mu)}{X(\mu)}\Big(1-\theta_+(\mu)\Big), \quad
-\infty<\mu<+\infty.
$$

This expression shows that for $\mu> 0$, the distribution
function for molecules
reflected from the wall exactly satisfies boundary condition (14).\\

\begin{center}
  \bf 5. CHEMICAL POTENTIAL JUMP AND PROFILE
\end{center}

Let us compare equalities (17) and (19). Using the second equality in (14),
we obtain the coefficient of the dimensionless chemical potential:
$$
 C(\alpha,q)=V_1(\alpha)-\dfrac{1-q}{q}
\dfrac{2l(\alpha)}{\ln(1-e^{\alpha})}.
\eqno{(20)}
$$

Figure 2 shows the dependence of coefficient $K(\alpha,q)$ on the
evapora\-ti\-on coefficient $q$;
curves 1 and 3 correspond to values of dimensionless che\-mi\-cal
potential $\alpha =-0.5$, and $-3$, respectively. The curve 2
correspods to case of Fermi--gases at $\alpha=-0.5$.

Figure 3 shows the dependence of coefficient $K(\alpha,q)$ on the
dimen\-si\-on\-less  chemical
potential; curves $1,2$ and $3$ correspond to the values of evaporation
coefficient $q=0.5, 0.3$ and $0.2$, respectively. The
curve 2 correspods to case of Fermi--gases at $q=0.5$.

Equality (17) describes the jump of the dimensionless chemical
poten\-tial. Passing in this equality to dimensional quantities,
we obtain the chemical potential jump in a quantum Bose gas:
$$
\delta \mu(0)=C(\alpha,q)l
\Big(\dfrac{d\mu(x_1)}{dx_1}\Big)_{x_1=+\infty},
$$
where $l=1/(\nu_1 \sqrt{\beta})$ is the mean free path of molecules.

We express collision frequency $\nu_1$ in accordance with relation (15).
Then the chemical potential jump (in terms of
dimensional quantities) can be calculated by the formula
$$
\delta \mu(0)=K(\alpha,q)\sqrt{\dfrac{m}{2kT}}D_{12}
\Big(\dfrac{d\mu(x_1)}{dx_1}\Big)_{x_1=+\infty},
$$
where $K(\alpha,q)$ is the coefficient of the chemical potential jump,
defined as
$$
K(\alpha,q)=\dfrac{2f_0(\alpha)}{l(\alpha)}
\Big[V_1(\alpha)-\dfrac{1-q}{q}
\dfrac{2l(\alpha)}{\ln(1-e^{\alpha})}\Big].
$$

Note that in the limit of high temperatures, when quantum
properties of the gas can be ignored, result (20) is transformed
into the following known
result for the concentration jump in a classical gas \cite{12}:
$$
C(-\infty,q)=1.0162+\sqrt{\pi}\,\dfrac{1-q}{q}.
$$

The chemical potential distribution in half-space
$x\geqslant 0$ (known as the chemical potential profile)
is defined by equality (7).
Substituting the expansion of distribution
(known as the chemical potential profile) is defined by equality (7).
Substituting the expansion of distribution function (16) into (7),
we find that the chemical potential profile can be constructed using
the formula
$$
\dfrac{a(x)}{G_\alpha}=C(\alpha,q)+x+\dfrac{1}{\pi}
\int\limits_{0}^{\infty}
\exp\Big(-\dfrac{x}{\eta}-V(\eta)\Big)\sin \zeta(\eta).
\eqno{(21)}
$$

\begin{center}
  \bf 6. CONCENTRATION JUMP AND PROFILE
\end{center}

Concentration profile  $N(x)$ of the gas in
half-space $x\geqslant 0$ is given by the equality
$$
N(x)=N_s+n(x),
$$
where
$$
N_s=\int f_F(C,\alpha)\,d\Omega, \qquad
n(x)=\int h(x,C_x)g(C,\alpha)\,d\Omega.
$$

We can easily find that $N_s=N_0l(\alpha)$, where
$$
N_0=\dfrac{2\pi (2s+1)m^3}{(2\pi \hbar)^3 (\sqrt{\beta})^3},
$$
and the deviation of the concentration
from concentration $n(x)$ of sa\-tu\-rated vapor is given by
$$
n(x)=P_N(x)G_N,
\eqno{(22)}
$$
\text{where}
$$
P_N(x)=\Bigg[C(\alpha,q)+x+\dfrac{1}{\pi}\int\limits_{0}^{\infty}
\exp\Big(-\dfrac{x}{\eta}-V(\eta)\Big)
\sin \zeta(\eta)\Bigg]g_2(\alpha),
$$
where $P_N(x)$ is the coefficient of the concentration profile,
$G_N$ is the quantity of the concentration gradient
(with respect to the dimensionless coordinate),
$$
G_N=\Big(\dfrac{dN}{dx}\Big)_{x=+\infty}.
$$

For the constant   gas   concentration,   the   chemical   potential
increases upon cooling \cite{LL}, so that $\alpha \to +\infty$
for  $T\to 0$, while for $T\to \infty$, $\alpha \to -\infty$.

Consequently, the curves plotted for small values of $\alpha$ correspond
to a lower effective gas temperature.

It follows from Eqs. (21) and (22) that the profiles of chemical
potential and concentration are proportional to within the
corresponding gradients.

In addition, it can easily be seen
that the gradients of chemical potential and concentration are
proportional to each other with a factor  $N_0$: $G_N=N_0G_\alpha$.

The latter equality indicates that the chemical potential
gradient is the gradient of the logarithm of concentration:
$$
G_\alpha=\Big(\dfrac{d\ln N}{dx}\Big)_{x=+\infty}.
$$

We can easily find that the relation between the deviations
of con\-cen\-tra\-tion and
chemical potential from the corresponding values at the wall is
$$
\dfrac{n(x)}{N_0}= g_2(\alpha) a(x).
\eqno{(23)}
$$

Using this relation, we can find the concentration jump for $x=0$,
$$
n_{as}(0)=N_0 g_2(\alpha) a_{as}(0)=C(\alpha,q)g_2(\alpha)G_N=
C_N(\alpha,q)G_N,
\eqno{(24)}
$$
where $C_N(\alpha,q)=C(\alpha,q)g_2(\alpha)$ is the coefficient of the
concentration jump.

The behavior of the concentration jump coefficient from relation (24)
is depicted in Figs. 4 and 5.

Figure 4 shows the dependence of the concentration jump coefficient
of the evaporation coefficient;
curves $1$ and $3$ correspond to dimen\-si\-on\-less chemical potential
$\alpha = -0.5$ and  $\alpha=-3$, respectively. The curve 2
corresponds to case of Fermi--gases at $\alpha=-0.5$.

Figure 5 shows the dependence of the
concentration jump coefficient on the chemical potential $\alpha$; curves
$1,3$, and $4$ correspond to values of accommodation coefficients
$q = 0.5, 0.3$, and $0.2$. The curve 2
corresponds to case of Fermi--gases at $q=0.5$.\\

\begin{center}
  \bf 7. CONCLUSIONS
\end{center}

The problem of evaporation of one of the binary gas components has
been solved analytically. The evaporating component is a Bose gas.
We studied the dependence of the chemical potential jump coefficient
on the evaporation coefficient and on the chemical potential.
On the basis of the analytic solution of the problem, an explicit
representation is obtained for the distribution function, as well
as the expressions for the chemical potential jump and the chemical
potential distribution in the half-space (chemical potential profile).
It is also shown that the concentration jump for the Bose gas and its
distribution in the half-space
are proportional to the chemical potential jump and distribution.
On Figs.  2-5 graph comparison of the received results with
similar results from work \cite{Kost} is shown .

\begin{figure}[h]
\begin{center}
\includegraphics[width=14cm,height=9cm]{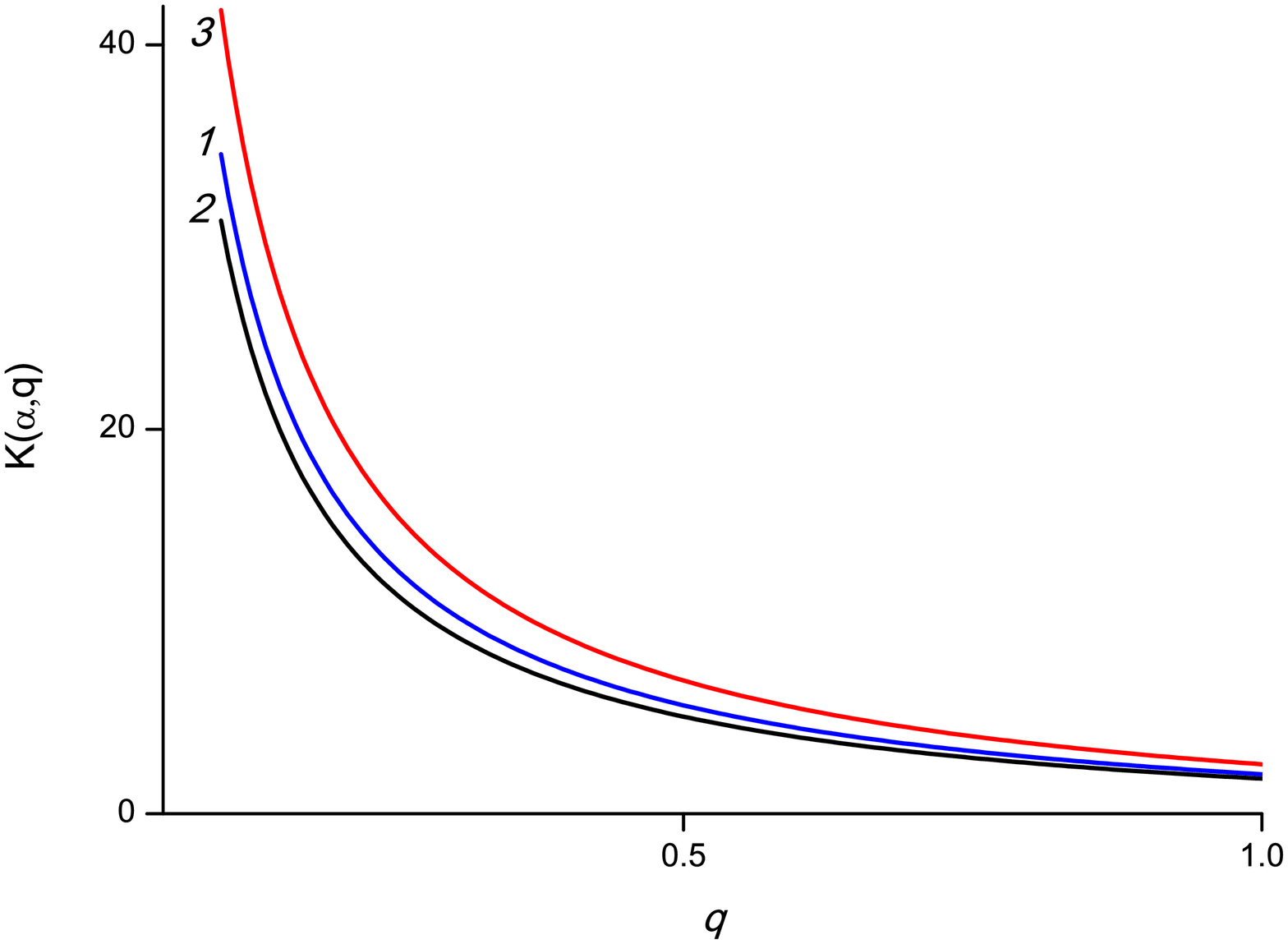}
\end{center}
\begin{center}
\caption{The dependence of coefficient $K(\alpha,q)$ on the
evaporation coefficient $q$;
curves 1 and 3 correspond to values of dimensionless chemical
potential $\alpha =-0.5$, and $-3$, respectively. The curve 2
correspods to case of Fermi--gases at $\alpha=-0.5$.}
\end{center}
%\end{figure}
%\begin{figure}[b]
\begin{center}
\includegraphics[width=14cm,height=9cm]{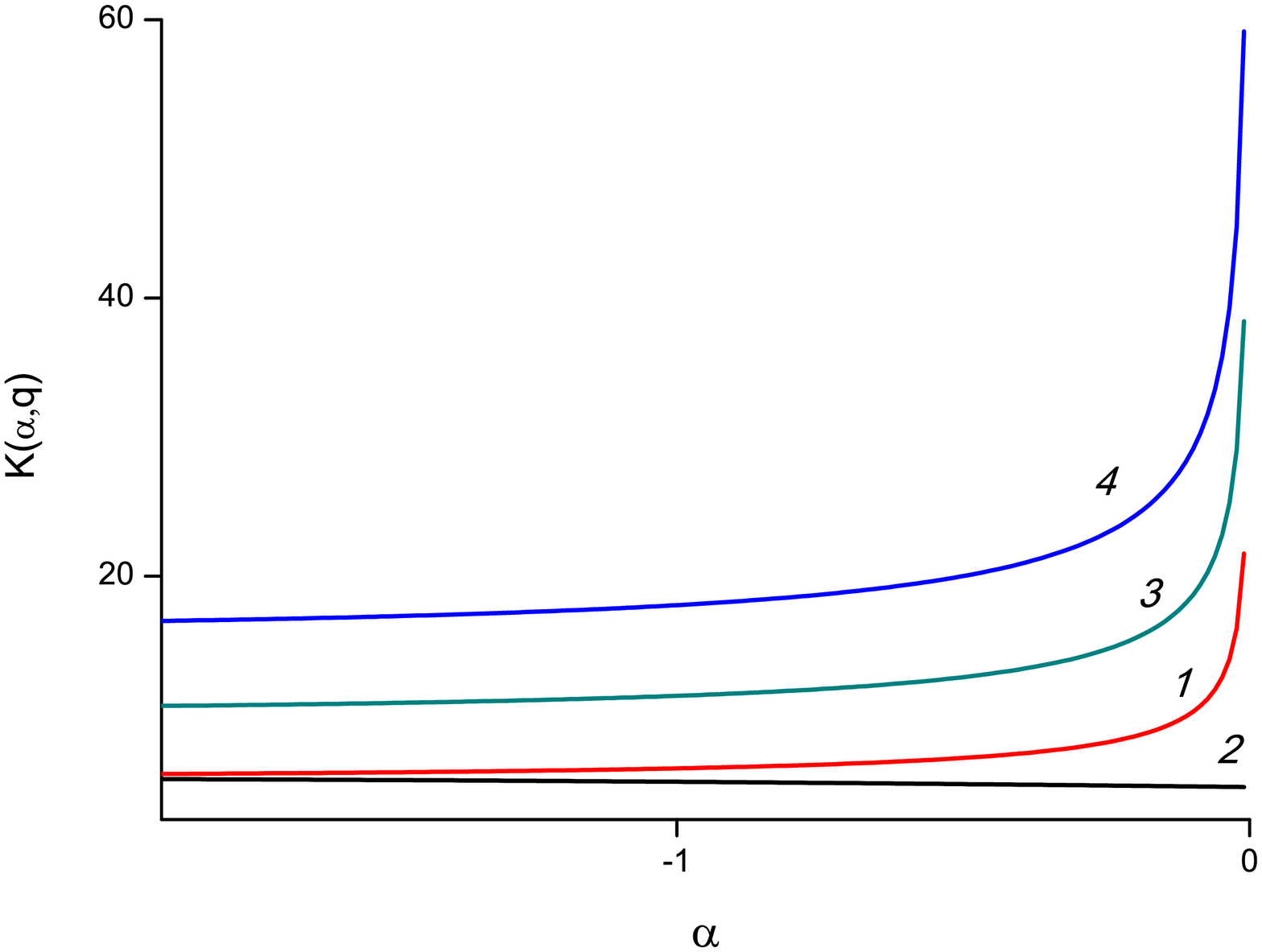}
\end{center}
\begin{center}
\caption{The dependence of coefficient $K(\alpha,q)$ on the
dimensionless  chemical
potential; curves $1,2$ and $3$ correspond to the values of evaporation
coefficient $q=0.5, 0.3$ and $0.2$, respectively. The
curve 2 correspods to case of Fermi--gases at $q=0.5$.}
\end{center}
\end{figure}

\begin{figure}[b]
\begin{center}
\includegraphics[width=14cm,height=9cm]{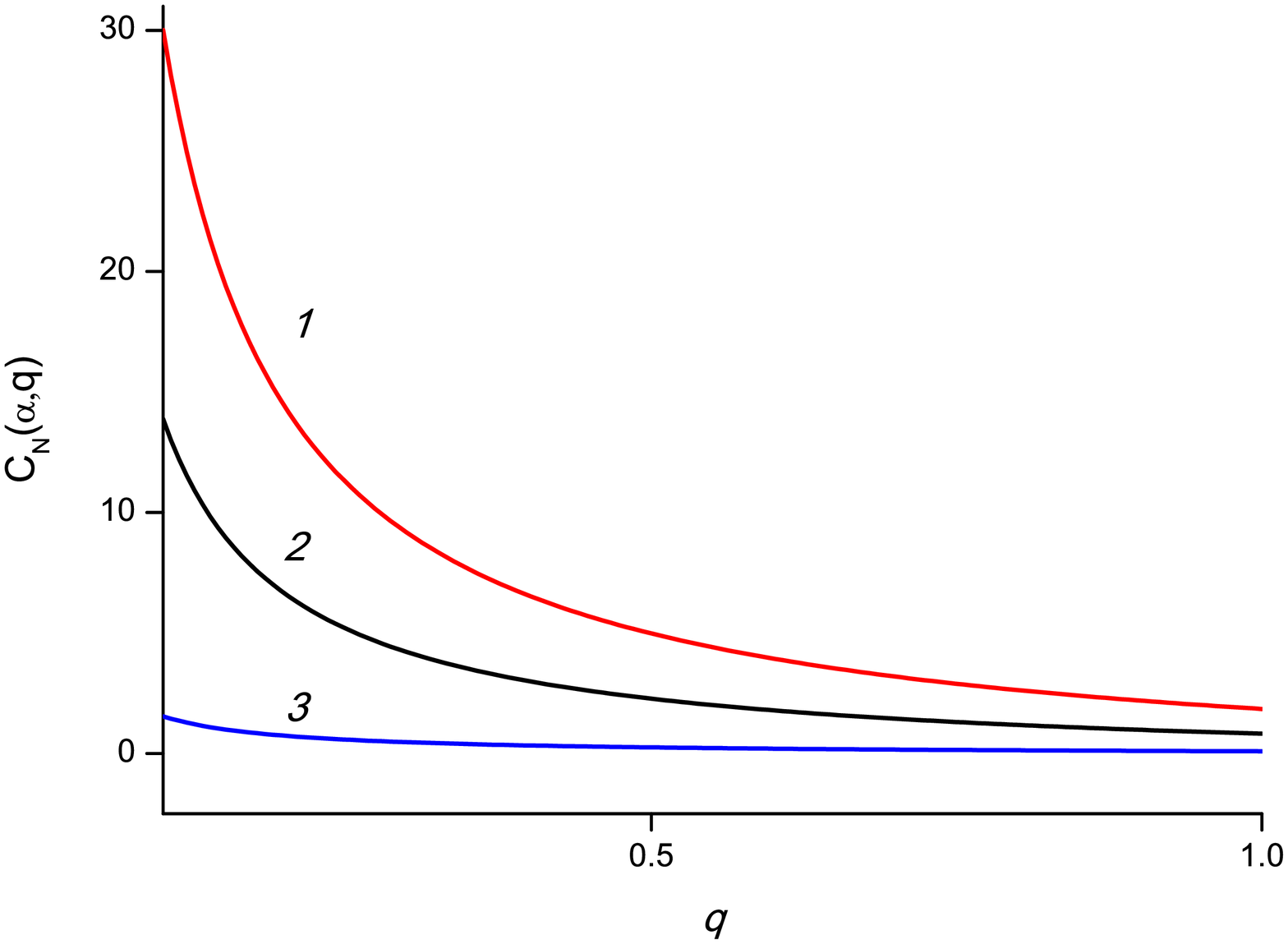}
\end{center}
\begin{center}
\caption{the dependence of the concentration jump coefficient
of the evaporation coefficient;
curves $1$ and $3$ correspond to dimen\-si\-on\-less chemical potential
$\alpha = -0.5$ and  $\alpha=-3$, respectively. The curve 2
corresponds to case of Fermi--gases at $\alpha=-0.5$.}
\end{center}
%\end{figure}
%\begin{figure}[t]
\begin{center}
\includegraphics[width=14cm,height=9cm]{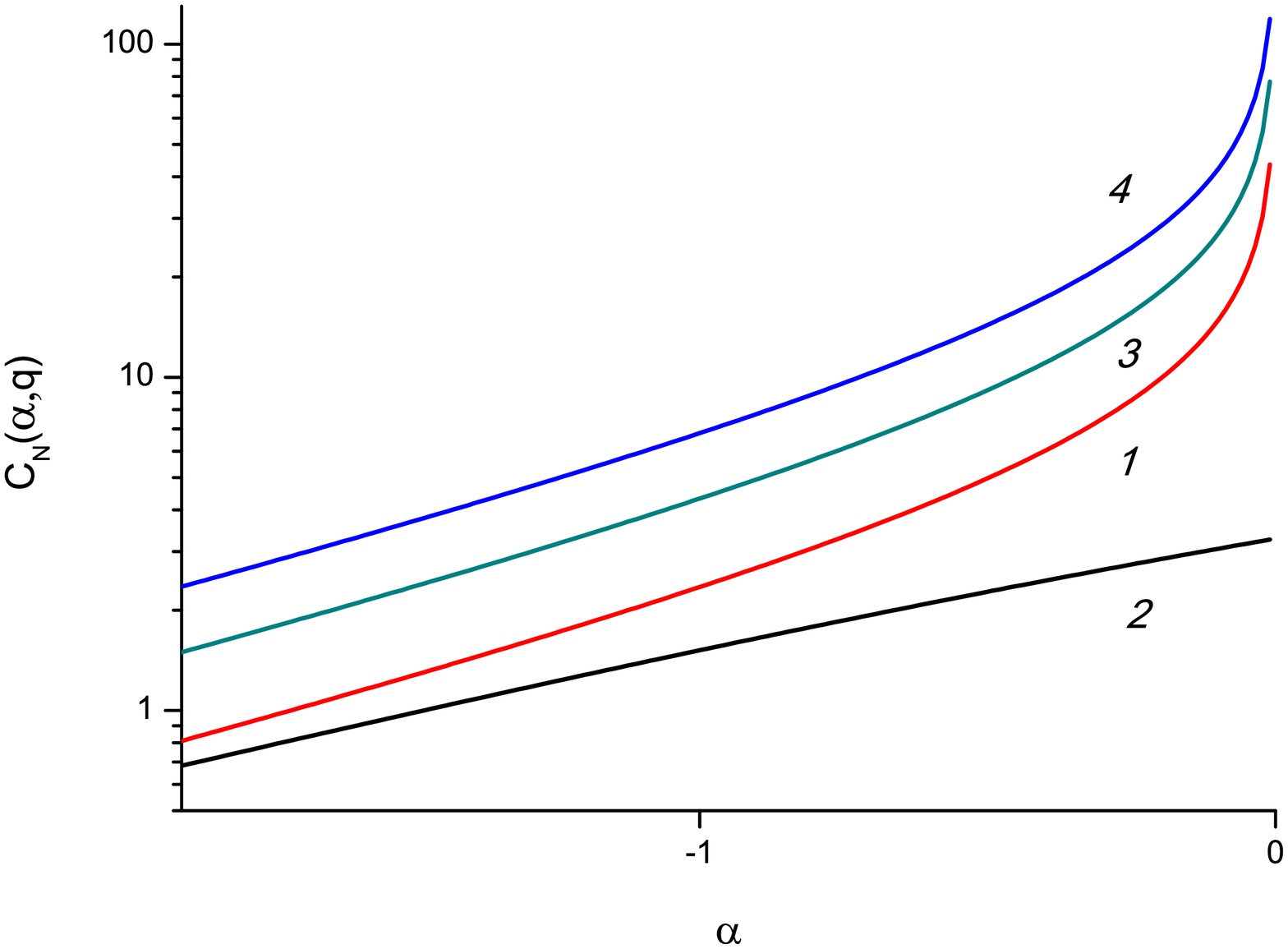}
\end{center}
\begin{center}
\caption{the dependence of the
concentration jump coefficient on the chemical potential $\alpha$; curves
$1,3$, and $4$ correspond to values of accommodation coefficients
$q = 0.5, 0.3$, and $0.2$. The curve 2
corresponds to case of Fermi--gases at $q=0.5$.}
\end{center}
\end{figure}

\clearpage
%\vspace{-2cm}

\end{document}